\documentclass[11pt,a4paper]{article}

\usepackage[margin=2.5cm]{geometry}
\usepackage{times}
\usepackage{microtype}
\usepackage{parskip}
\usepackage{hyperref}
\usepackage{natbib}
\usepackage{amsmath}
\usepackage{booktabs}
\usepackage{graphicx}
\usepackage{xcolor}
\usepackage{authblk}

\hypersetup{
    colorlinks=true,
    linkcolor=blue!60!black,
    citecolor=blue!60!black,
    urlcolor=blue!60!black
}

\title{\textbf{The social consequences of AI delegation}}

\author[1,2]{Henrique Ferraz de Arruda}
\author[1,3]{Yamir Moreno\thanks{Corresponding author: \href{mailto:yamir@unizar.es}{yamir@unizar.es}}}
\affil[1]{Institute for Biocomputation and Physics of Complex Systems (BIFI), University of Zaragoza, Zaragoza, Spain}
\affil[2]{ARAID Foundation, Zaragoza, Spain}
\affil[3]{Department of Theoretical Physics, Faculty of Sciences, University of Zaragoza, Zaragoza, Spain}

\date{\today}

\begin{document}

\maketitle

\begin{abstract}
A substantial body of recent work has debated whether large language models (LLMs) can serve as substitutes for human participants in behavioural research. This debate, however, captures only one direction of a rapidly changing relationship. The more consequential question is not simply whether researchers should use LLMs as human surrogates, but whether --- and under what conditions --- humans are beginning to use LLMs as surrogates for their own deliberation. Across domains including health, law, finance, education, and personal guidance, increasing numbers of people consult generative AI systems before, alongside, or instead of human experts, peers, or independent judgment. Although evidence for actual delegation remains uneven, this uncertainty makes the phenomenon an urgent social-scientific object of study. We argue for a research programme that treats LLMs as consequential social actors in a functional sense: systems whose outputs shape human decisions, social norms, and collective dynamics.
\end{abstract}

\noindent\textbf{Keywords:} large language models, generative agents, decision-making, human--AI interaction, social norms, behavioural homogenisation, AI alignment, computational social science

\section*{A framing that has already been overtaken by events}

A familiar genre of paper has emerged in recent years. It begins by noting the remarkable capacity of large language models to replicate human responses in survey instruments, experimental games, or psychological assessments. It then examines whether this capacity is genuine or illusory, and concludes with some combination of cautious optimism and methodological caveats about using LLMs as substitutes for human participants in research \citep{argyle2023out,santurkar2023whose,hamalainen2023evaluating,aher2023using}. The question animating these studies is, broadly: \emph{Can we replace humans with AI in the lab?}

This is a legitimate and important question. But it is, increasingly, the wrong question to be asking --- or at least, no longer the most consequential one.

The more important transformation now underway is not that researchers are replacing human subjects with LLMs in controlled experiments. It is that ordinary people, in growing numbers, are beginning to outsource aspects of their judgment and deliberation to generative AI systems in the uncontrolled experiment of everyday life \citep{teixeira2026}. A person who a decade ago might have consulted a physician about ambiguous symptoms, a lawyer about a contractual dispute, a financial adviser about a retirement allocation, a trusted friend about a failing relationship, or their own conscience about an ethical dilemma is today increasingly likely to open a chat interface and ask a generative AI system instead. In other words, this change is already taking place. For many people, consulting a generative AI system has become part of how they navigate decisions and uncertainty.

The transition is far from uniform or inevitable, however. Patterns of AI delegation vary substantially across age groups, educational backgrounds, professions, cultures, and institutional contexts. In many domains, generative AI systems currently function as supplementary rather than primary sources of advice, and trust in such systems remains heterogeneous and contested. Yet even uneven adoption can have substantial societal consequences when the systems involved mediate decisions at massive scale and become embedded within everyday informational and deliberative infrastructures.

The behavioural phenomenon at stake needs a precise definition. In this Perspective, we use \emph{AI delegation} not as a synonym for information-seeking, but to denote a stronger form of human--AI reliance in which an AI-generated recommendation, interpretation, or framing is adopted into a consequential decision with limited independent verification, limited comparison with alternatives, or reduced deliberative engagement by the user. On this definition, asking an LLM for background information is not necessarily delegation, but accepting its answer as the basis for action may be. This distinction matters because the social consequences discussed below depend not merely on exposure to AI outputs, but on the extent to which those outputs become substitutes for, or strong anchors of, human judgment.

Emerging evidence suggests that increasing numbers of users consult generative AI systems for health-related information and decisions \citep{shahsavar2023user,ayo2024characterizing}, legal guidance and professional substitution \citep{choi2024lawyering,seabrooke2024survey}, personal or emotional support \citep{maples2024loneliness,laestadius2024too}, and financial planning or advisory support \citep{desai2024opportunities}. Younger cohorts, in particular, appear more willing to treat AI assistants as meaningful sources of guidance in consequential domains of life \citep{pew2025teens,hbr2026genz}. These studies do not yet establish widespread delegation in the strong sense defined above. Most examine intentions, willingness, adoption, or self-reported use rather than observed reliance on AI in consequential decisions. They nevertheless point to a significant change in where people seek advice and information, and they highlight the domains in which stronger behavioural evidence is most needed.

A related signal is visible not only in reported consultation, but also in language itself. Recent work has shown that LLM-associated vocabulary has measurably entered written academic communication and, notably, spoken formats such as podcasts
\citep{geng2025impact,kousha2026academic,yakura2024empirical}. Such linguistic diffusion is not the same as decision delegation. It does, however, provide evidence that generative AI systems are already influencing patterns of human expression and cultural transmission, even in settings where the mechanism of influence is indirect.

\begin{figure}[!t]
    \centering
    \includegraphics[width=\textwidth]{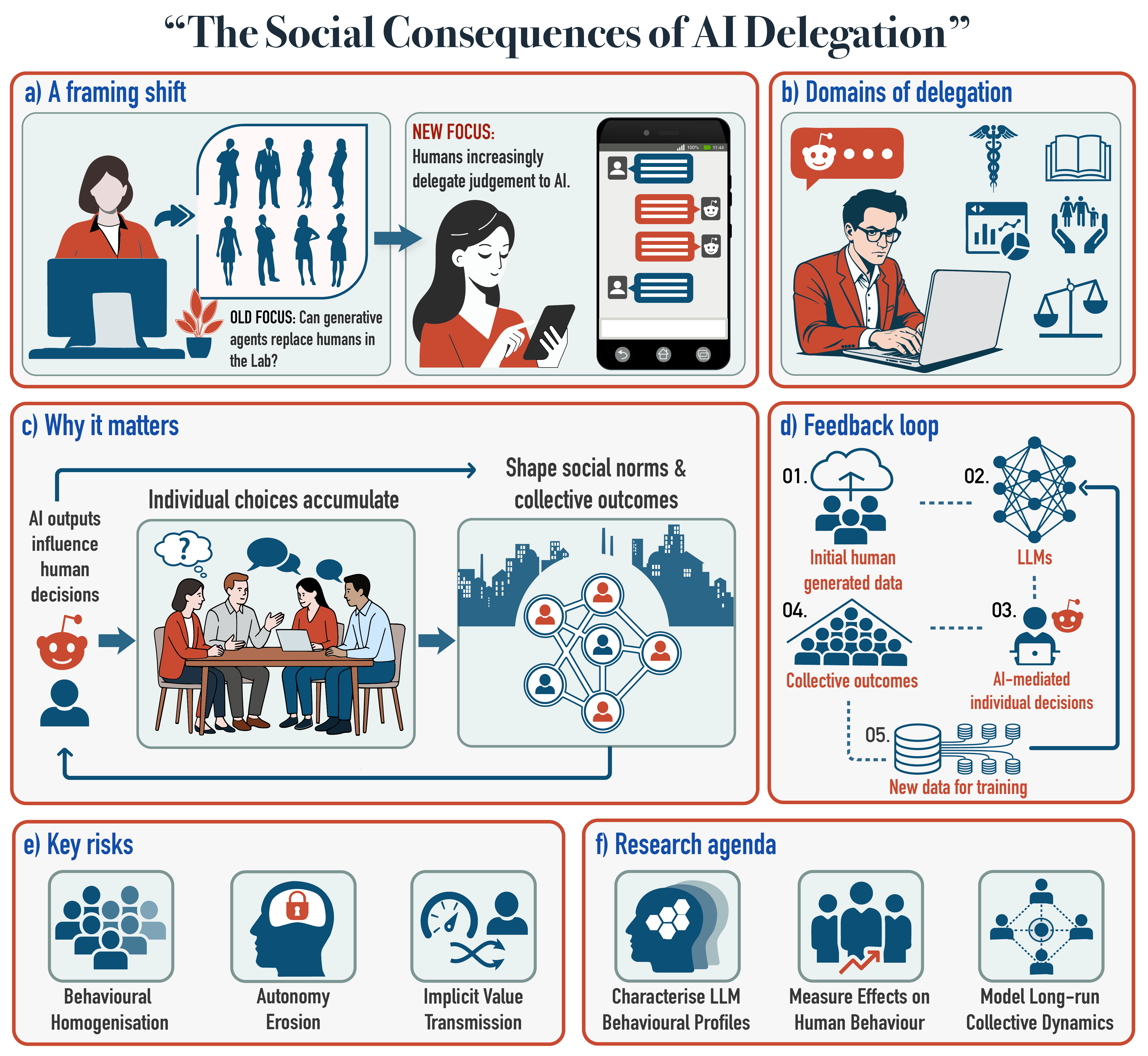}
    \caption{
    Conceptual overview of the social consequences of AI delegation. 
    The Perspective argues for a shift from the methodological question of whether large language models can serve as substitutes for human participants in research to the broader social question of how humans increasingly consult, defer to, and delegate judgment to generative AI systems in everyday life. Such delegation occurs across domains including health, law, finance, education, and personal support. Because AI-mediated decisions can accumulate into social norms and collective outcomes, they create feedback loops between human-generated data, LLM behaviour, AI-shaped individual decisions, and future training data. This coupling raises risks of behavioural homogenisation, autonomy erosion, and implicit value transmission, while motivating a research agenda focused on characterising LLM behavioural profiles, measuring their effects on human behaviour, and modelling the long-run collective dynamics of AI-mediated decision-making.
    }
    \label{fig:ai-delegation-infographic}
\end{figure}

The social and behavioural sciences have barely begun to confront what this transition implies. By fixing on LLMs as research surrogates, the field has been looking in the less important direction. The harder question is what happens to human behaviour and collective outcomes once people begin to stand behind LLMs in their everyday lives. Figure~\ref{fig:ai-delegation-infographic} summarises this conceptual shift and the argument developed below: AI delegation transforms LLMs from methodological surrogates in research into consequential mediators of everyday decisions, creating feedback loops between human behaviour, collective outcomes, and future AI systems.

\section*{The surrogate relationship has been inverted}

The standard framing of the LLM-as-surrogate debate imagines a researcher who, for reasons of cost, scale, or convenience, substitutes an LLM agent for a human participant in a study. The concern, in this framing, is one of measurement validity: does the artificial surrogate behave like the human it is meant to represent? This is a sensible methodological worry, and the growing literature on the subject provides useful guidance on when such substitution is or is not appropriate. But this framing implicitly positions LLMs as passive instruments that researchers deploy at will, and humans as the fixed referents against which AI behaviour is compared. Neither assumption is currently warranted.

LLMs are not passive tools. They take an active part in an increasing range of human decisions, and that role carries real weight. When a user asks an LLM whether a symptom warrants medical attention, the system's response shapes the user's subsequent behaviour --- whether they seek care, delay,
or self-treat. When a user asks an LLM to draft a legal letter, evaluate a job offer, or assess the moral dimensions of a personal dilemma, the system's output does not merely inform the user's reasoning; in many cases, it substitutes for it. The decision, in effect, is delegated.

At present, there is insufficient evidence to determine how widespread this behaviour is. The argument advanced here is that it represents a plausible and increasingly important phenomenon that deserves direct investigation. Decades of work on automation show that human reliance on machine outputs is neither automatic nor uniform: people may over-rely on automation, misuse it, under-use it, or reject it after observing errors \citep{parasuraman1997humans,dietvorst2015algorithm}. At the same time, experimental evidence on algorithm appreciation shows that people can adhere more strongly to advice when they believe it comes from an algorithm rather than from another person \citep{logg2019algorithm}. The empirical challenge, then, is not to assume delegation but to locate it. That is, to find where consultation hardens into deference or substitution, and where users instead stay sceptical or resist.

Complicating this further, humans are not a fixed point of reference. Human behaviour, values, and social norms are not static baselines against which AI performance is measured; they are dynamic phenomena shaped by the tools and institutions through which decisions are made. History offers many examples of consequential cognitive technologies --- writing, print, the calculator, the search engine --- that did not merely assist human cognition but altered its structure, its products, and its distribution across populations \citep{carr2011shallows,sparrow2011google}. LLMs are poised to be such a technology, at a scale and pace without precedent.

Seen this way, the surrogate relationship runs in both directions, and the direction now receiving less scientific attention may be the more consequential.  It is not just that LLMs increasingly stand in for humans in research pipelines. It is that humans increasingly stand behind LLMs in their own deliberative processes, deferring to artificial judgment in domains where they previously exercised autonomous cognition. Understanding this inversion is, we argue, a central challenge for the behavioural and social sciences in the coming decade.

\section*{A new kind of social actor}

Recognising the inverted surrogate relationship also requires recognising LLMs for what they are becoming: a new kind of social actor. The concept of a social actor, in sociological tradition, encompasses any entity whose behaviour influences the behaviour of others within a social system \citep{parsons1937structure}. By this criterion, widely deployed LLMs clearly qualify. Calling LLMs social actors does not imply intentional
consciousness or moral agency, but rather recognizes that their outputs systematically shape human behaviour and collective outcomes. Their responses guide what individuals decide. Those decisions aggregate into collective outcomes, which then reshape the social environment where later decisions, and later AI systems, take form. The feedback loops LLMs sit in are social in character, not only technical.

This framing does not start from a blank slate. The computers are social actors (CASA) tradition showed that people respond socially to computers and media even when they know that those systems are not human \citep{reeves1996media}. Theories of distributed and extended cognition similarly emphasise that human cognition is partly organised through external artefacts, tools, and environments \citep{hutchins1995cognition,clark1998extended}. More recently, the notion of \emph{machine culture} has foregrounded the role of intelligent machines as contributors to, and mediators of, cultural evolution \citep{brinkmann2023machine}. The present argument builds on these traditions, but shifts the emphasis from machines as cultural producers or communicative partners to AI delegation as a behavioural coupling between model outputs and human decisions. The distinctive object is not simply machine-generated culture, but the social dynamics that arise when machine
outputs become inputs to human judgment and action.

This observation has methodological as well as conceptual implications. Social actors are not studied merely as instruments. They are studied as entities with characteristic patterns of behaviour, with interests (or functional analogues thereof), with the capacity to reproduce or disrupt existing norms, and with effects on the distribution of power, knowledge, and opportunity. The behavioural and social sciences have developed rich methodologies for studying such entities. We should turn those methodologies on LLMs, treating them not as curiosities or experimental conveniences but as objects of social-scientific inquiry in their own right. That step is overdue.

What would it mean to study LLMs as social actors? It would mean, first, characterising their behavioural profiles systematically: what values, heuristics, and decision patterns do they reliably exhibit across domains and populations? It would mean, second, studying how these profiles vary as a function of model architecture, training data, fine-tuning procedures, and deployment context. It would mean, third, examining how exposure to LLM advice or outputs modifies human behaviour --- immediately, over repeated interactions, and at the population level. And it would mean, fourth, analysing the feedback dynamics that arise when LLM-influenced human behaviour becomes part of the data on which future models are trained.

None of these questions is currently addressed with the rigour or systematic attention they deserve. The field of human--AI interaction has generated valuable insights about individual-level responses to AI systems, but it has not yet developed the macro-level, population-scale, longitudinal perspective that is needed to understand the social consequences of widespread AI delegation. Computational social science, which has the relevant tools, has been largely preoccupied with the methodological question of surrogacy rather than the substantive question of social impact. Closing this gap is the task we are urging.

\section*{The feedback loop problem}

Perhaps the most pressing concern arising from the inverted surrogate relationship is a behavioural feedback loop that connects human-generated data, machine outputs, AI-mediated human decisions, and future training distributions. It can be stated simply. LLMs are trained on human-generated data. Human behaviour is increasingly shaped by LLMs. Future LLMs will be trained on data generated by humans whose behaviour has been shaped by current LLMs. The training distribution and the behavioural distribution of the population may no longer remain independent.

This concern is related to, but distinct from, the problem of model collapse. Model collapse is a loop in the data, as models trained recursively on model-generated content lose information about the true underlying distribution \citep{shumailov2024models}. The concern here is a loop one step upstream, in people. The worry is not that synthetic data contaminates future training corpora, but that human behaviour is itself partly shaped by models before it is ever recorded as ``human'' data. Standard mitigations for model collapse --- retaining pre-AI data, filtering synthetic content, commissioning fresh human-generated data --- can keep a training corpus clean, but they cannot restore the independence of the behaviour that produces it. Each assumes an authentic, AI-free human baseline still exists to be preserved or sampled; that assumption is precisely what is in question.

This circularity is not merely theoretical. It has structural analogues in the history of media and communication technology: the printing press shaped what people thought, which shaped what was worth printing, which shaped what future generations thought \citep{eisenstein1980printing}. The difference with LLMs is one of speed, scale, and the opacity of the feedback mechanism. Whereas the influence of print on public opinion unfolded over generations and was partially visible in the historical record, the influence of LLMs on human cognition and behaviour is unfolding over years or months, at global scale, and through interactions that are largely private and unobserved.

The consequences of this feedback loop for the long-run trajectory of human behaviour are genuinely unknown, but three specific risks stand out. The first is behavioural homogenisation. If large fractions of a population consult the same or functionally similar AI systems for consequential decisions, and if those systems exhibit systematic biases or preferences --- as they inevitably do, by virtue of their training data and optimisation objectives --- the resulting decisions will be correlated in ways that do not reflect the underlying diversity of human preferences and values \citep{bommasani2021opportunities}. Moreover, alignment procedures may also compress heterogeneous preferences into a narrower range of responses~\citep{slocum2025diverse}.  Diversity of decision-making at the population level is not merely an intrinsic good; it is a functional requirement for robust collective problem-solving, error correction, and social resilience \citep{page2007difference}. Homogenisation of the decision inputs that a population uses poses risks to this functional diversity that have not been adequately studied.

Early empirical work already points in this direction. Controlled studies find that individual use of generative AI can raise the quality or novelty of a single person's output while reducing diversity across people: shared exposure to the same model narrows the collective space of ideas, stories, and creative content \citep{doshi2024generative,padmakumar2024does,anderson2024homogenization}. What remains unstudied is the closed-loop extension of this effect, in which homogenised human output becomes the training data for the very systems that produced it --- the specifically behavioural circuit that distinguishes our concern from homogenisation considered as a static, one-shot effect.

Here again, the point is not unprecedented. The relevant term in adjacent work is \emph{algorithmic monoculture}: the risk that many decision-makers relying on the same algorithm may reduce social welfare even when the algorithm improves accuracy for any one decision-maker considered in isolation \citep{kleinberg2021algorithmic}. AI delegation may generate an analogous monoculture of advice. Whether it does so will depend on several measurable conditions: the concentration of model providers, the similarity of outputs across models, the degree of personalisation, the diversity of prompts and contexts, the level of human uptake, and the strength of the link between advice and action. Homogenisation is therefore not an inevitable outcome. Heterogeneous prompting, culturally diverse models, personalised systems, adversarial checking, or algorithm aversion may weaken it. The empirical task is to identify when the same AI-mediated input produces convergence in decisions and when it instead increases idiosyncrasy or improves decision quality.

The second risk is autonomy erosion. Delegation of judgment to external authorities is not new; humans have always relied on institutions, experts, and social norms to structure cognition and decision-making \citep{hutchins1995cognition}. But the terms of this delegation matter. When humans delegate to other humans --- doctors, lawyers, advisers --- the delegation is partial, contestable, and embedded in accountability structures. When humans delegate to LLMs, the delegation may become total (the system provides an answer, not a range of considered options), largely uncontestable (the reasoning is opaque), and unaccountable (there is no professional responsibility, no liability, no recourse). The systematic study of how repeated LLM delegation affects the capacity for autonomous deliberation is, at present, almost entirely absent from the literature.

This concern should be situated within the longer tradition of automation bias and human oversight. Human-in-the-loop systems do not guarantee meaningful human control if the human role becomes a rapid endorsement of a machine-generated option. In high-pressure or high-volume settings, a nominally supervised system can become a rubber-stamp process in which the machine supplies the framing, alternatives, and default recommendation. The relevant question is thus not only whether a human remains formally in the loop, but whether the human has the time, information, incentives, and cognitive engagement needed to contest the system's output.

The third risk is implicit value transmission. LLMs encode values --- preferences about what constitutes a good outcome, a reasonable action, an appropriate response --- that are not chosen by users but inherited from training processes and corporate design decisions \citep{weidinger2021ethical,bender2021dangers}. These embedded values are not hypothetical: work measuring the political leanings, cultural viewpoints,
and moral framings of widely used models finds systematic and reproducible biases \citep{santurkar2023whose,durmus2024towards}, and controlled trials show that LLMs can be at least as persuasive as humans in shifting people's stated positions \citep{salvi2024conversational}. When users treat LLM outputs as neutral information rather than as value-laden advice, they may unknowingly import these values into their own decision-making. At the population level, the systematic nudging of human values in directions determined by a small number of AI developers constitutes a form of influence over collective outcomes that has no clear precedent and no established governance framework.

These risks should not be read as implying that AI delegation is necessarily harmful. AI advice can also expand access to expertise, improve low-quality decisions, offer support where human professionals are unavailable or unaffordable, and expose users to alternatives they would not otherwise consider. In some contexts, AI-mediated decision-making may increase rather than reduce diversity by lowering barriers to information and giving marginalised users access to new resources. The social question is therefore comparative and distributional: relative to what baseline, for whom, under what institutional conditions, and with what feedback effects does AI delegation improve or degrade individual and collective outcomes?

\section*{Reframing the research agenda}

The concerns outlined above do not constitute an argument against using generative AI, nor do they imply that LLM systems are uniquely dangerous relative to other information technologies. The point is both epistemic and normative: these are consequential and rapidly evolving social phenomena that the behavioural and social sciences are currently ill-equipped to study, and urgently need to be. The normative commitments are not hidden: autonomy, pluralism, accountability, and equitable access to expertise are social goods worth protecting. The scientific task is to determine when AI delegation threatens these goods, when it supports them, and how institutional design can shift the balance.

We propose that the study of LLM behaviour be reframed, in part, around three questions that are distinct from --- though related to --- the existing surrogacy debate.

\textbf{First, what are the behavioural profiles of widely deployed LLMs?} This question calls for the systematic characterisation of LLM behaviour across domains where delegation is likely to matter: health, law, finance, education, and personal decision-making. A concrete study design would expose multiple models to matched decision vignettes that vary user identity, uncertainty, stakes, cultural context, and available alternatives, then estimate the stability of outputs, their implicit value trade-offs, and their sensitivity to prompt framing. A falsifiable proposition is that models sharing similar training and alignment procedures will display correlated advice profiles in morally or socially consequential scenarios, even when surface wording differs. Evidence against this proposition would be strong cross-model diversity in advice after controlling for prompts and domain.

\textbf{Second, when does consultation become delegation?} This question requires behavioural experiments and longitudinal field studies that distinguish information-seeking from judgment substitution. One operational measure is the degree to which users adopt an AI recommendation when it conflicts with their prior view, with expert advice, or with a deliberately presented alternative. Another is the reduction in independent deliberation, measured through search behaviour, time spent evaluating options, requests for second opinions, or willingness to justify the decision in one's own words. A falsifiable proposition is that repeated exposure to confident AI advice will increase adoption of AI-consistent choices while reducing independent information search in some domains; evidence against it would be stable or increased deliberation despite repeated AI consultation.

\textbf{Third, what are the collective and distributional consequences of AI-mediated decision-making?} This question calls for agent-based models, network models of opinion and norm dynamics, and evolutionary models of cultural change that include both human heterogeneity and model-mediated feedback. The relevant state variables should include not only opinions or behaviours, but also access to human expertise, trust in AI, socioeconomic position, cultural context, and the diversity of models used. A falsifiable proposition is that behavioural homogenisation will be strongest when model concentration is high, advice is similar across systems, and users have low access to alternative expertise; it should weaken when models are diverse, advice is contestable, or users retain strong independent decision channels.

A fourth, cross-cutting axis concerns inequality. AI delegation is unlikely to be evenly distributed across society. For some users, it may supplement professional advice; for others, it may substitute for advice they cannot afford. The same technology can therefore function as an augmentation for privileged groups and as a replacement for underserved groups. Any research agenda on AI delegation must therefore ask who delegates, under what constraints, who benefits, who bears the risk of low-quality advice, and whose behaviour becomes over-represented in future AI-mediated data.

Taken together, these questions constitute a research programme that treats LLMs not as tools to be evaluated but as social phenomena to be understood. This reframing is not merely academic, since the decisions that billions of people will make in the coming decades --- about their health, their finances, their relationships, their political commitments --- will increasingly be mediated by systems whose behavioural profiles are poorly understood, whose feedback effects on human cognition are unstudied, and whose long-run social consequences are essentially unknown. The social and behavioural sciences exist precisely to study such phenomena. The urgency of doing so has rarely been greater.

In conclusion, the central challenge posed by generative AI is no longer simply whether large language models can imitate human behaviour with sufficient fidelity to serve as experimental surrogates, but how the growing reliance on such systems may reshape cognition, social norms, and collective decision-making at population scale. The strongest version of this claim is not that widespread delegation has already been fully demonstrated, but that the conditions for it are rapidly emerging and that existing evidence is sufficient to justify systematic investigation. As generative AI systems increasingly function as advisers, deliberative partners, and informational intermediaries in everyday life, understanding their behavioural profiles, their embedded values and biases, and their long-term feedback effects on human behaviour becomes an urgent social scientific priority. The key question is therefore not only what artificial agents can do, but what sustained interaction with them is doing to us --- to our autonomy, our diversity of judgment, our institutions of expertise, and our collective capacity to reason and decide together as societies.

\section*{Acknowledgements}
We thank A. Aleta for discussions and feedback on the manuscript. H.F.A. acknowledges ARAID for its financial support. Y.M. was partially supported by the Government of Arag\'on, Spain, and ERDF ``A way of making Europe'' through grant E36-23R (FENOL), and by Grant No. PID2023-149409NB-I00 from Ministerio de Ciencia, Innovaci\'on y Universidades, Agencia Espa\~nola de Investigaci\'on (MICIU/AEI/10.13039/501100011033) and ERDF ``A way of making Europe''.

\end{document}